\documentclass[screen, nonacm]{acmart}

 \usepackage{soul}
\usepackage[table]{xcolor}
\usepackage{tabularx}   
\usepackage{booktabs}   
\usepackage{array}      
  \definecolor{DarkGreen}{HTML}{008000}

   \AtBeginDocument{%
  }
     \setcopyright{acmlicensed}
\copyrightyear{2026}
\acmYear{2026}
\acmDOI{}

\keywords{Generative AI, reading, self-regulated learning, cognitive engagement, college students, human-AI interaction}
          \acmISBN{978-1-4503-XXXX-X/18/06}

\begin{document}
    \title[How College Students Use AI to Navigate Course Readings]{How College Students Use AI to Navigate Course Readings: Evidence from an Eight-Week Study}

 \begin{abstract} 
College students increasingly use AI chatbots to support academic reading, yet we lack granular understanding of how these interactions shape their reading experience and cognitive engagement. We conducted an eight-week longitudinal study with 15 undergraduates who used AI to support assigned readings in a course. We collected 838 prompts across 239 reading sessions and developed a coding schema categorizing prompts into four cognitive themes: Decoding, Comprehension, Reasoning, and Metacognition. Comprehension prompts dominated (59.6\%), with Reasoning (29.8\%), Metacognition (8.5\%), and Decoding (2.1\%) less frequent. Most sessions (72\%) contained exactly three prompts, the required minimum of the reading assignment. Within sessions, students showed natural cognitive progression from comprehension toward reasoning, but this progression was truncated. Across eight weeks, students' engagement patterns remained stable, with substantial individual differences persisting throughout. Qualitative analysis revealed an intention-behavior gap: students recognized that effective prompting required effort but rarely applied this knowledge, with efficiency emerging as the primary driver. Students also strategically triaged their engagement based on interest and academic pressures, exhibiting a novel pattern of \textit{reading through AI} rather than with it: using AI-generated summaries as primary material to filter which sections merited deeper attention. We discuss design implications for AI reading systems that scaffold sustained cognitive engagement.

 \end{abstract}
                  
\author{Yue Fu}
\authornote{Corresponding author.}
\email{chrisfu@uw.edu}
\orcid{0000-0001-5828-5932}
\affiliation{%
   \institution{University of Washington}
  \city{Seattle}
  \state{Washington}
  \country{USA}
}

\author{Joel Wester}
\email{joel.wester@di.ku.dk}
\orcid{0000-0001-6332-9493}
\affiliation{%
  \institution{University of Copenhagen}
  \city{Copenhagen}
  \country{Denmark}
}

\author{Niels van Berkel}
\email{nielsvanberkel@cs.aau.dk}
\orcid{0000-0001-5106-7692}
\affiliation{%
  \institution{Aalborg University}
  \city{Aalborg}
  \country{Denmark}
}

\author{Alexis Hiniker}
\email{alexisr@uw.edu}
\orcid{0000-0003-1607-0778}
\affiliation{%
  \institution{University of Washington}
  \city{Seattle}
  \state{Washington}
  \country{USA}
}
 \maketitle
   
\section{Introduction}
      
Reading is foundational for cognitive development and shapes people's understanding of the world. Students' transition to higher education marks a critical juncture in this process: reading demands increase significantly~\cite{holschuh2019college}, and students are required to engage with complex textbooks, research articles, and technical reports, often with minimal guidance. 
Research indicates that over 80\% of college tasks involve reading, requiring not only comprehension but also high-level analysis, interpretation, and the integration of multiple sources~\cite{holschuh2019college}. Yet, many students struggle with these expectations, facing anxiety and difficulties connecting reading with writing~\cite{bergman2024students}. While effective reading requires independent strategies that combine metacognitive and cognitive skills~\cite{nist2002college}, balancing necessary scaffolding with the goal of fostering student independence remains a critical challenge~\cite{holschuh2018comprehension}.

Recently, college students have widely adopted generative AI chatbots to support their reading tasks. A recent Australian survey of undergraduate philosophy students reports that over half use chatbot-based AI tools to engage with course readings.~\cite{corbin2024reading}. These tools, including specialized interfaces like \textit{ChatPDF}\footnote{https://www.chatpdf.com/} and \textit{ChatDOC}\footnote{https://chatdoc.com/}, streamline the reading process by answering comprehension questions, defining terms, and generating summaries~\cite{etkin2024differential}.
However, the integration of AI into reading practices yields mixed outcomes regarding cognitive development. On one hand, AI tools offer personalized learning and simplify complex texts, potentially enhancing reading effectiveness for some learners~\cite{zheng2024exploration, ademola2024reading}. On the other hand, reliance on AI may prioritize convenience over critical evaluation, promoting passive reading behaviors~\cite{cahyani2023ai, ademola2024reading} and undermining deep cognitive engagement. Furthermore, the impact of AI appears non-uniform: lower-performing students often benefit through gains in comprehension and test performance, while higher-performing peers may experience declines~\cite{etkin2024differential}. Empirical research remains in an early stage and lacks a granular understanding of students' interactions with AI during self-regulated reading tasks.

A understanding of how students use AI in their reading and the corresponding cognitive involvement could open new opportunities for systems designers to develop AI systems that better support reading and learning while discouraging counterproductive engagement patterns. Furthermore, understanding students' AI use patterns in reading assignments can inform the design of self-monitoring and teacher-monitoring systems that support both students and teachers during self-reflection reading tasks. Lastly, understanding individual students' usage and behavior can inform building learner models that enable personalized tutoring systems to scaffold each student's distinct learning needs~\cite{mcnichols2025studychat}. 
 We conducted an eight-week longitudinal study in an undergraduate course to capture students' AI usage patterns for self-regulated reading. 
Specifically, we address the following research questions:

\begin{itemize}
    \item \textbf{RQ1:} How do college students use AI to support their reading tasks?
    \item \textbf{RQ2:} How do students' AI usage patterns and cognitive engagement evolve over time?
    \item \textbf{RQ3:} Why do students engage with AI in the ways they do?
\end{itemize}

We recruited 15 undergraduate students enrolled in an introductory AI course at a major US public university. Over eight weeks, students submitted interaction logs documenting their use of AI to support their assigned readings. We collected 838 prompts across 239 reading sessions. We developed a coding schema grounded in reading comprehension literature and the data, categorizing prompts into four cognitive themes: Decoding, Comprehension, Reasoning, and Metacognition. In addition, we conducted interviews with five participants to understand the motivations and perspectives behind their usage patterns.

Our quantitative analysis revealed that students primarily used AI for comprehension support (59.6\% of prompts), with reasoning (29.8\%) and metacognition (8.5\%) less frequent. Most sessions (72\%) contained exactly three prompts, which was the required minimum, and only 4.3\% of prompts incorporated prompt engineering techniques despite explicit instruction. Within sessions, students showed cognitive progression from comprehension toward reasoning, but this progression was truncated by limited engagement. Across eight weeks, usage patterns remained stable, with individual differences persisting throughout the study.

Our qualitative findings revealed there is an intention-behavior gap: students recognized that effective prompting required effort and yielded better results, yet they rarely applied this knowledge in practice. Efficiency emerged as the primary driver, with students strategically triaging their engagement based on interest and academic pressures. Students also exhibited a novel pattern of \textit{reading through AI} rather than \textit{with} it, using AI-generated summaries as primary material and selectively engaging with original texts. Furthermore, students articulated tensions inherent in AI-supported reading, recognizing the benefits such as convenience, personalized response, and access to extended knowledge, but simultaneously risked shallow processing, diminished independent thinking, and incomplete understanding.

This paper makes the following contributions:
\begin{itemize}
    \item An empirical characterization of how college students use AI chatbots for class reading, including a coding schema of four cognitive themes and related ten codes.
    \item Evidence that students exhibit limited engagement, with stable individual differences in usage patterns across an eight-week period.
    \item Identification of an intention-behavior gap where students' stated understanding of effective AI use diverges from their actual practice.
    \item Design implications for AI reading support systems that scaffold deeper cognitive engagement while accommodating students' efficiency-driven orientations.
\end{itemize}

\section{Related Work}

\subsection{Reading in College}

When students transition to college, the demands of academic reading increase dramatically~\cite{holschuh2019college}. College students are expected to comprehend complex textbooks, research journal articles, technical and professional materials, industry reports, and media. Over 80\% of college tasks involve reading, and these tasks typically require higher-order comprehension and the ability to integrate multiple sources~\cite{holschuh2019college}. These reading tasks require more than mere understanding; they require analysis, interpretation, comparison, and the application of knowledge from the reading materials. 

College reading is also inherently self-regulated~\cite{stang2016active, hollander2022importance, lepore2019read}. Instructors assign readings and expect students to complete them outside of class, often without direct instruction or immediate accountability. Students make strategic decisions about whether and how to read based on time constraints, grade implications, and perceived payoff~\cite{hollander2022importance}. The paradox is that while higher education relies heavily on reading, it provides little support for the new content, context, and forms of reading students encounter~\cite{lepore2019read}. Students face multiple challenges, including difficulty interpreting complex academic texts, overwhelming and anxiety-inducing experiences, a lack of explicit guidance on disciplinary reading practices, inconsistent expectations across courses, and struggles integrating reading with writing~\cite{bergman2024students}.
 
Recent research has begun examining AI-assisted reading experiences, though most studies focus on English as a Second Language (ESL) learners~\cite{pan2025effects}, whose reading materials differ substantially from the complex texts assigned in college courses. As students increasingly use AI for reading and other college assignments, questions remain: How do students use AI for college reading? And how can systems better support them? Our study addresses these questions by examining students' AI usage patterns during reading, discussing implications for learning and cognition, and offering design suggestions for future AI reading support systems.

\subsection{Reading Theories and Strategies}
 Academic reading requires building meaning across a text and integrating it with prior knowledge. Classic text comprehension models describe how readers build meaning both from individual sentence level (microstructure) and from the global meaning of the text (macrostructure)~\cite{Kintsch1978TowardAM}. Reading research distinguishes \emph{skills} from deliberate, goal-directed \emph{strategies}~\cite{afflerbach2008clarifying}. Skills are automatic actions that support decoding/comprehension with speed, efficiency, and fluency, usually without conscious awareness or deliberate control. On the other hand, reading strategies are deliberate, goal-directed attempts to control and modify the reader’s efforts during readings. Effective readers flexibly select strategies such as summarizing, questioning, clarifying, making inferences, and evaluating ideas in response to the text and task goals~\cite{afflerbach2009identifying,pressley2012verbal}. College reading materials require both skills to decode local meanings and strategies to construct and monitor broader understanding.

Self-regulated learning (SRL) theory frames these strategy choices as part of an active process in which learners set goals, monitor progress, and regulate cognition, motivation, and behavior under task constraints~\cite{pintrich2000role, zimmerman2002becoming}. In reading contexts, self-regulated reading includes planning how to approach a text, monitoring comprehension during reading, and evaluating progress relative to task demands and reading purposes~\cite{hu2017using}. Researchers have used various methods to study self-regulated reading, including interviews~\cite{fadlelmula2010learner}, observations~\cite{veenman2005relation}, questionnaires~\cite{liyanage2012gender}, and particularly think-aloud protocols~\cite{hu2017using, pressley2012verbal, braaten2003longitudinal}, which allow researchers to access readers' cognition and metacognition in the self-regulation process in the moment. 

SRL theory further argues that learning typically produces analyzable ``observable traces'' (e.g., notes, summaries, self-generated questions) that externalize regulation~\cite{inbook}. When students read with the help of AI, the prompts they provide to AI systems can serve as a parallel trace. In our study, we position users' prompts to AI as externalized trace data of questioning, monitoring, and help-seeking. To characterize the cognitive engagement reflected in these traces, we draw on learning cognition frameworks including Bloom's taxonomy~\cite{bloom1956taxonomy, adams2015bloom}, metacognition research~\cite{veenman2006metacognition}, reading comprehension taxonomies such as Barrett's taxonomy~\cite{barrett1968barrett}, and Constructive Responsive Reading Strategies~\cite{pressley2012verbal, afflerbach2009identifying, afflerbach2008clarifying}. We combined these theoretical foundations with patterns observed in our data to develop a coding schema for analyzing users' prompts to AI (see Section \ref{coding_schema} for method details).

\subsection{AI-Supported Reading and Its Effect}
                           
HCI research has long explored how technology can support reading. Early systems focused on augmenting specific aspects of the reading process, such as supporting comprehension through analogies~\cite{shao2025studentunderstanding} or metaphors~\cite{yarmand2025metaphors}. Broader platforms like Semantic Reader~\cite{lo2023semanticreaderprojectaugmenting}, Elicit~\cite{whitfield2023elicit}, and Marvista~\cite{chen2023marvista} offered diverse ways of interacting with text. More recently, the emergence of large language models has spurred a new wave of reading support tools that enable conversational interaction with texts (e.g.,~\cite{zhu2025autopbl, watanabe2025echoread, xu2025revisemate}). These systems go beyond static augmentation by allowing students to ask questions, request summaries, and explore ideas through dialogue with AI. As these tools gain adoption, a growing body of empirical work has begun evaluating their effects on reading and learning.

Several studies report promising outcomes. Etkin et al.\ compared different forms of AI reading support, including AI-generated summaries, outlines, and chatbots, and found that these tools positively influenced lower-performing students, though higher-performing students may experience declines~\cite{etkin2024differential}. Prajapati and Das shows integrating AI-enabled tools within traditional learning contexts can influence how students engage with complex texts \cite{prajapati2025activereading}. They found that participants using a smartphone companion application with AI-based question-answering features actively leveraged the AI to search for information and clarify their understanding during academic reading tasks, often alternating between printed textbook content and AI responses before revisiting the source material for verification. Pan et al.\ examined a GenAI-powered chatbot used as a reading companion in out-of-class reading, drawing on self-determination theory. Students with chatbot access demonstrated significantly higher autonomous reading motivation and engagement, though no significant differences in reading performance were observed~\cite{pan2025readingcompanion}. Taken together, these findings suggest that AI tools can enhance motivation and accessibility. 
                
However, other studies highlight important challenges. Zheng and Fan reported mixed outcomes for AI-supported reading, with benefits accompanied by potential costs for learning. In their quasi-experiment, the AI-assisted group showed significantly lower concentration, and participant feedback suggested tool-related distractions, such as uncertainty about answer accuracy and slow response generation, can interrupt sustained attention. The authors also note weaker performance in retelling background/purpose and a tendency toward lower recall of methodological and detailed information,~\cite{zheng2024exploration}. In another study, Zhang et al.\ found that students often guessed, felt overwhelmed, or developed workarounds when AI output did not align with their expectations~\cite{zhang2025navigatingthefog}. Beyond cognitive concerns, researchers have identified technical, social and institutional challenges. These include students' tendencies to use AI secretly~\cite{zhang2025secretuse, fu2026everyone} or to avoid disclosing their use~\cite{adnin2025studentperspectives}, as well as concerns from both teachers and students about shifting classroom dynamics, deskilling of instructors, and difficulties meeting individual needs~\cite{doyle2025readtheirwork}. 

While this body of work has advanced our understanding of AI-supported reading tools and their effects, most studies evaluate specific tool interventions or measure outcomes at a single time point. Less is known about how students' own patterns of reading with AI tools of their choosing develop over time and shape cognitive engagement. Our study addresses this gap by examining students' AI use for reading across an eight-week period.
         
\section{Method}

\subsection{Course Context and Participants}

We recruited participants from an introductory AI elective course offered during the winter quarter of 2025 at a large public R1 university in the United States. The 10-week course introduced students to AI concepts through both theoretical and practical components. Each week featured two sessions: (a) a two-hour interactive lecture exploring a specific AI-related topic, complemented by in-class activities and reflective discussions on societal and personal implications, and (b) a two-hour studio session focused on hands-on, project-based exploration of AI tools. The course aimed to equip students with the skills to work effectively with AI in their future studies and careers.

For weekly reading assignments, students chose between two submission formats: a traditional written reflection (200--500 words) or an AI-supported reading log (described in the next subsection). The instructor presented both options during the first class session; all students opted for the AI-supported format.

Students were informed that research participation was entirely voluntary and would not affect their course grades. Students received compensation for AI subscription costs (up to \$40) for the duration of the course, regardless of participation. Of the 21 enrolled students, 18 consented to participate in the study. And after the exclusion (see detail in Section \ref{dataset}), 15 participants were included in the study. Participants were not restricted to specific AI platforms, though the instructor suggested popular tools such as ChatGPT, Gemini, Perplexity, and Claude. Based on the instructor's observations and discussions with students, most participants used these suggested tools. Five students subsequently participated in an interview after the class was concluded, and they were compensated \$40 per hour. Participants' demographic data are presented in Table ~\ref{table:participant_table}.

\begin{table}[htbp]
\centering
\caption{Participant Demographics and AI Tool Experience}
\label{tab:demographics}
\small
\begin{tabular}{@{}clll@{}}
\toprule
\textbf{ID} & \textbf{Year} & \textbf{AI Tools Used} & \textbf{AI Experience} \\
\midrule
P1  & Senior    & ChatGPT                                      & Since Jan 2022 \\
P2  & Junior    & ChatGPT, Grammarly GO                        & Since 2022 \\
P3  & Freshman  & ChatGPT, Gemini, Perplexity, Quillbot        & Since Aug 2021 \\
P4  & Freshman  & ChatGPT, Quillbot, Canva                     & Since Jan 2023 \\
P5  & Junior    & ChatGPT, Canva                               & Since Mar 2023 \\
P6  & Sophomore & ChatGPT, Gemini                              & Since 2022 \\
P7  & Senior    & ChatGPT, Claude, Perplexity                  & Since Jan 2023 \\
P8  & Junior    & ChatGPT                                      & Since Winter 2023 \\
P9  & Senior    & ChatGPT                                      & Since Sep 2023 \\
P10 & Senior    & ChatGPT, Canva                               & Since 2023 \\
P11 & Senior    & ChatGPT, Claude                              & Since Mar 2024 \\
P12 & Junior    & ChatGPT, Gemini, Claude, Copilot, + 4 others & Since Jan 2023 \\
P13 & Junior    & ChatGPT, Gemini, Grammarly GO, Canva         & Since 2023 \\
P14 & Junior    & ChatGPT, Gemini                              & Since Feb 2023 \\
P15 & Junior    & ChatGPT                                      & Since 2024 \\
\bottomrule
\end{tabular}
\vspace{1mm}
\begin{minipage}{0.9\linewidth}
\footnotesize
\textit{Note.} AI Experience indicates when participants first started using AI chatbot tools. P12's additional tools include Grok, Grammarly GO, Quillbot, and Canva.
\label{table:participant_table}
\end{minipage}
\end{table}
  
\begin{table}[ht]
\centering
\caption{Example reading session from one participant. User prompts are shown in full; AI responses and reflections are truncated for brevity.}
\label{tab:example_session}
\small
\renewcommand{\arraystretch}{1.2}
\definecolor{AIbg}{gray}{0.92}
\begin{tabularx}{\textwidth}{@{}>{\raggedright\arraybackslash}p{0.18\textwidth} X@{}}
\toprule
\textbf{Component} & \textbf{Content} \\
\midrule
\multicolumn{2}{@{}l}{\textit{Conversation with AI}} \\
\addlinespace[0.3em]

User Prompt 1 & Summarize this reading, giving me the key points, analyzing the argument and making insightful conclusions \\
\rowcolor{AIbg}
AI Response 1 & Certainly, here's a summary of the reading. Key Points: 2024 as ``Primordial Soup'': The article positions 2024 as a year of intense activity and rapid development in the AI field\ldots \\
\addlinespace[0.2em]

User Prompt 2 & Provide this in a more digestible format \\
\rowcolor{AIbg}
AI Response 2 & Think of 2024 as the ``wild west'' of AI---tons of new ideas, but a bit chaotic\ldots \\
\addlinespace[0.2em]

User Prompt 3 & Expand on the contents of the article \\
\rowcolor{AIbg}
AI Response 3 & Here's a deeper dive into the key themes\ldots \\

\midrule
\multicolumn{2}{@{}l}{\textit{Participant Reflections (Optional)}} \\
\addlinespace[0.3em]

On AI Interaction & The initial prompt was reasonably useful, however it missed several key details. The second prompt resulted in shorter sentences and more accessible language, but focused more on conciseness rather than conveying all information. \\
\addlinespace[0.2em]

On the Reading & It was interesting to see how this article approached AI through a business lens, focusing less on the overall industry\ldots \\

\midrule
\multicolumn{2}{@{}l}{\textit{Discussion Questions for Class}} \\
\addlinespace[0.3em]

& 1. What are potential benefits and drawbacks to each approach to AI development? \newline
2. How can we maintain responsibility and ethics in the AI age? \newline
3. What are implications of AI search becoming a key form of information retrieval? \\

\bottomrule
\end{tabularx}
\end{table}

\subsection{Researcher Positionality}
One author served as the course instructor, which afforded deep contextual knowledge of the learning environment but also introduced potential concerns. To mitigate power dynamics, we emphasized to students that research participation was voluntary and independent of course performance. The course instructor interviewed one participant, while research assistants who were not affiliated with the course conducted the remaining interviews. All data were anonymized prior to coding. We acknowledge that social desirability effects may have influenced student behavior. Students might have presented more favorable accounts of their AI engagement during interviews or crafted their reading logs with awareness that submissions would be reviewed after the course was concluded. To minimize these effects, we encouraged honest responses during interviews and conducted all interviews after final grades had been posted. The study was approved by our institution's Institutional Review Board.

\subsection{Procedure}

Each week, the course assigned two required readings and several optional readings on AI-related topics, including AI for creativity, AI-supported communication, AI's influence on the workforce, and AI companions. Reading materials included academic papers, industry reports, blog posts, book chapters, and news articles, typically ranging from 5--20 pages, with occasional longer industry reports (up to 68 pages including appendices).

Students submitted reading logs from Week 1 through Week 8 via Google Forms. Each log documented a single reading session and contained: (1) the complete history of prompts submitted to the AI, (2) the AI's responses, (3) an optional reflection on the reading material, (4) an optional reflection on interacting with AI, and (5) discussion questions for in-class use (see Table~\ref{tab:example_session} for an example). Data collection began in Week 1, following the institution's add/drop period. To account for readings from Week 0, Week 1 required four submissions; subsequent weeks required a minimum of two entries.

The reading assignment task page provided students with general suggestions for using AI to support sense-making, such as seeking clarification on complex points, analyzing implications of the work, and drawing connections to other knowledge. The learning goals were twofold: to familiarize students with upcoming course topics and to generate discussion questions for class. Each reading session required a minimum of three conversational turns (one user prompt followed by one AI response), with no upper limit. Students were instructed to continue interacting until they reached a satisfactory level of understanding. The grading policy emphasized good-faith engagement: students received credit for submitting assignments, with exceptional reflections or creative AI use earning additional points. The teacher assistant reviewed submissions weekly and collaborate with the instructor to incorporate selected student-generated questions into class discussions.

During the first week, the instructor delivered a lecture on prompt engineering using the CO-STAR framework~\cite{costar, teo_2023} (Context, Objective, Style, Tone, Audience, and Response format) and reinforced these techniques throughout the course. Thus, students were familiar with prompt engineering strategies from the outset.

\subsection{Dataset}
\label{dataset}
We collected 290 reading session entries from 18 consenting participants. During initial data cleaning, we excluded three participants who consistently demonstrated minimal engagement, operationalized as: (a) using only generic very short prompts across all sessions, or (b) submitting templated copy-and-paste prompts identical across sessions. The final dataset comprised 239 complete session entries from 15 participants, containing a total of 838 individual prompts. Each session entry included the participant's prompts, the AI's responses, optional reflections on both the reading and the AI interaction, and proposed discussion questions for class.

\subsection{Coding Schema}
\label{coding_schema}
To analyze students' prompting strategies, we developed a coding schema informed by established frameworks on reading comprehension and cognitive engagement. Two researchers reviewed the dataset alongside relevant literature, including the Revised Bloom's Taxonomy~\cite{adams2015bloom}, Constructive Responsive Reading Strategies~\cite{pressley2012verbal, afflerbach2009identifying, afflerbach2008clarifying}, Barrett's Taxonomy of Reading Comprehension~\cite{clymer1968reading, barrett1968barrett}, and metacognition and learning frameworks~\cite{veenman2006metacognition}. We also consulted recent codebooks and datasets for analyzing student prompts to AI chatbots~\cite{prasad2024self, ma2025scaffolding, mcnichols2025studychat}.

One researcher first reviewed prompts and AI responses from 100 randomly selected entries, generating initial codes. The goal was to characterize students' usage patterns and the cognitive demands reflected in their prompts. Two researchers then discussed these initial codes, refined definitions, and independently applied them to an additional 100 entries. The two researchers met regularly to discuss discrepancies, refine code definitions, and iterate on the coding book. To address our research questions regarding cognitive engagement, we aggregated the codes into four higher-level themes: Decoding, Comprehension, Reasoning, and Metacognition (see definition in Table \ref{tab:themes}), grounded in reading comprehension and metacognition literature~\cite{afflerbach2008clarifying, veenman2006metacognition}. We also found a small percentage of prompts were difficult to code since 1) semantic meaning is not clear 2) too much grammar mistakes and broken sentences hard to interpret 3) intent of the prompt is vague or too broad. We coded them as ``unknown''. To establish reliability, the two researchers then independently coded another randomized 100 entries, and inter-rater agreement was calculated using Cohen’s Kappa. After confirming reaching the interrater reliability (Cohen's $k$ = 0.824) and the codebook (see Table \ref{tab:prompt_codes}), the two researchers divided the remaining dataset and applied the codebook to all users' prompts. The two researchers met again to discuss the unknown prompts and, after reaching consensus, recoded them into other codes. If, after discussion, the unknown prompts could not be agreed upon between two researchers, we leave them as ``unknown''. Lastly, one researcher reviewed all the coding.

\subsection{Data Analysis}
We analyzed two types of data: (1) students' prompts to AI, which we coded using the schema described in Section \ref{coding_schema}, and (2) qualitative data from interviews and optional reflections on AI usage. While each session's AI responses were recorded and reviewed to provide context for interpreting students' prompts, we treated the prompts themselves as the primary unit of analysis, aligning with our framing of prompts as externalized traces of cognitive engagement. A systematic evaluation of the quality, depth, or accuracy of the AI responses was beyond the scope of this study. We suggest analysis of AI responses as a direction for future work in Section~\ref{limitation and future work}.
 
\subsubsection{Quantitative and Qualitative Analysis of Students' AI Prompts} To examine patterns in students' AI usage, we analyzed the distribution of codes and cognitive themes across 239 reading sessions, prompt positions, and weeks. We used chi-square tests to examine associations between categorical variables (e.g., prompt position and cognitive theme), linear mixed-effects models to account for repeated measures within participants, and intraclass correlation coefficients (ICCs) to quantify the extent to which engagement patterns reflected stable individual differences versus within-person variation. Complementing this statistical approach, we also performed a qualitative content analysis of the prompts to interpret the quantitative results. We reviewed the specific text of prompts associated with key statistical trends to identify the context and student intent driving the observed patterns.

\subsubsection{Qualitative Analysis of Interview and Students' Optional Reflections on AI Usage}
We analyzed interview transcripts and students' optional reflections on AI usage using thematic analysis~\cite{braun2019reflecting, braun2021thematic}. We collected 109 reflections across the 239 reading sessions. One author first reviewed all interview transcripts and reflection entries, generating initial codes to capture patterns in students' motivations, strategies, and perceived benefits and challenges of AI-supported reading. A second author reviewed the data and discussed the codes together. Themes emerged from the discussion and one author organized them with representative quotes for subsequent research team discussion. Through several rounds of discussion with the entire research team, the team refined the themes and resolved discrepancies. One author then consolidated the themes and further extracted quotes. Interview and reflection data were analyzed together, as both sources provided complementary perspectives on students' reasoning behind observed usage patterns: interviews offered retrospective accounts of general practices, while reflections captured in-the-moment responses to reading sessions.

\begin{table}[htbp]
\centering
\caption{Cognitive Themes Used to Classify Students' AI Prompts}
\label{tab:themes}
\small
\renewcommand{\arraystretch}{1.4}
\begin{tabular}{p{0.15\textwidth} p{0.52\textwidth} p{0.25\textwidth}}
\toprule
\textbf{Theme} & \textbf{Definition} & \textbf{Codes} \\
\midrule

Decoding & 
Prompts focused on surface-level clarification, defining terms, and repairing local understanding. & 
\texttt{VOCAB} \\
\addlinespace[0.4em]

Comprehension & 
Prompts focused on building a global understanding of the text's explicit meaning and structure. & 
\texttt{LOC}, \texttt{SUMM}, \texttt{EXPL} \\
\addlinespace[0.4em]

Reasoning & 
Prompts requiring the student to go beyond the text to infer, apply, or compare ideas. & 
\texttt{INFER}, \texttt{APPLY}, \texttt{COMP} \\
\addlinespace[0.4em]

Metacognition & 
Prompts focused on self-regulation, where the student monitors their own understanding, checks their knowledge, and evaluate the reading. & 
\texttt{MON}, \texttt{EVAL} \\

\bottomrule
\end{tabular}
\end{table}

\begin{table}[htbp]
    \centering
    \small
    \renewcommand{\arraystretch}{1.5}
      \caption{Prompt Codebook with Definitions and Example Prompts}
    \begin{tabularx}{\textwidth}{
        >{\raggedright\arraybackslash\hsize=0.7\hsize\bfseries}X
        >{\raggedright\arraybackslash\hsize=0.9\hsize}X
        >{\raggedright\arraybackslash\hsize=1.4\hsize}X
    }
    \toprule
    \textbf{Code} \newline \normalfont\textit{\small(High-Level Theme)} & \textbf{Definition} & \textbf{Example Prompts} \\
    \midrule

    Vocabulary term (\texttt{VOCAB}) \newline \normalfont\textit{\small(Decoding)} &
    Requests the meaning of a word or phrase \emph{as used in the text} (i.e., local, in-context definition). &
    ``What does \dots{} mean in this sentence?'' \newline
    ``What does \dots{} refer to?'' \\

    \midrule 

    Locating (\texttt{LOC}) \newline \normalfont\textit{\small(Comprehension)} &
    Prompt asks \textit{where} the text says something, where a term is defined, or \textit{whether} the text mentions something. &
    ``Where does the article say ... ?'' \newline
    ``Does the reading discusses \dots{}?'' \newline
    ``Does the reading mention \dots{} ?'' \\

    Summary (\texttt{SUMM}) \newline \normalfont\textit{\small(Comprehension)} &
    Requests an overview of the main ideas, key points, or takeaways of the text or a section. &
    ``Can you summarize the main ideas of this article?'' \newline
    ``Give me takeaways.'' \\

    Explanation (\texttt{EXPL}) \newline \normalfont\textit{\small(Comprehension)} &
    Requests clarification of concepts, ideas, arguments in detail; requesting backgrounds, contexts, examples or ask for a simpler rephrasing to support understanding. &
    ``Give me more background on ... ?'' \newline
    ``What is the purpose of this section \dots{})?'' \newline
    ``Rephrase this paragraph in simpler terms for me to understand.'' \\

    \midrule 

    Inference (\texttt{INFER}) \newline \normalfont\textit{\small(Reasoning)} &
    Draws conclusions beyond what is explicitly stated---e.g., identifying assumptions, implications, or downstream consequences. &
    ``What assumptions are being made here?'' \newline
    ``What are the implications of these results for \dots{}?'' \newline
    ``What does this suggest about how \dots{} might change in the future?'' \\

    Application (\texttt{APPLY}) \newline \normalfont\textit{\small(Reasoning)} &
    Applies ideas from the reading to a new case, domain, or the student’s own experience. &
    ``How would this framework apply to ... ?'' \newline
    ``How can I use these findings in my own work or life as a STEM student?'' \\

    Comparison (\texttt{COMP}) \newline \normalfont\textit{\small(Reasoning)} &
    Compares or contrasts two cases, timeframes, contexts, often asking for similarities/differences. &
    ``Compare how this would affect education versus entertainment.'' \newline
    ``How is this technology compared to the historical discover of \dots{}?'' \\

    \midrule 

    Evaluation (\texttt{EVAL}) \newline \normalfont\textit{\small(Metacognition)} &
    Evaluates the credibility, quality, evidence, or methodology of the reading (e.g., limitations, validity, credibility). &
    ``The article mentioned ... but is it true .... ?'' \newline
    ``Critique the methodology?'' \\

    Monitoring understanding (\texttt{MON}) \newline \normalfont\textit{\small(Metacognition)} &
    Checks or verifies one’s understanding, or signals confusion and asks for help diagnosing it. &
    ``My understanding is that \dots{} is that correct?'' \newline
    ``The article suggests ... , but I think ... . Does that make sense?'' \\

    \midrule 

    Unknown (\texttt{UNKNOWN}) \newline \normalfont\textit{\small(Unknown)} &
    The prompt cannot be reliably interpreted or coded (e.g., unclear intent or copy-pasted text without a question). &
    [Copies and pastes an sentence from the article with no question or instruction.] \\

    \bottomrule
    \end{tabularx}
  
    \label{tab:prompt_codes}
\end{table}

\section{Students' Prompt Analysis Results}
\label{results:72_percent}
To address RQ1 (How do college students use AI to support their self-regulated reading tasks?) and RQ2 (How do students’ AI usage patterns and cognitive engagement evolve over time?), we analyzed prompts from 15 participants across 239 reading sessions that spanned eight weeks. Participants generated 838 total prompts, of which 822 were reliably coded into cognitive themes (16 prompts, or 1.9\%, were classified as Unknown due to unclear intent). 

\begin{figure}[htbp]
    \centering
    \includegraphics[width=0.60\textwidth]{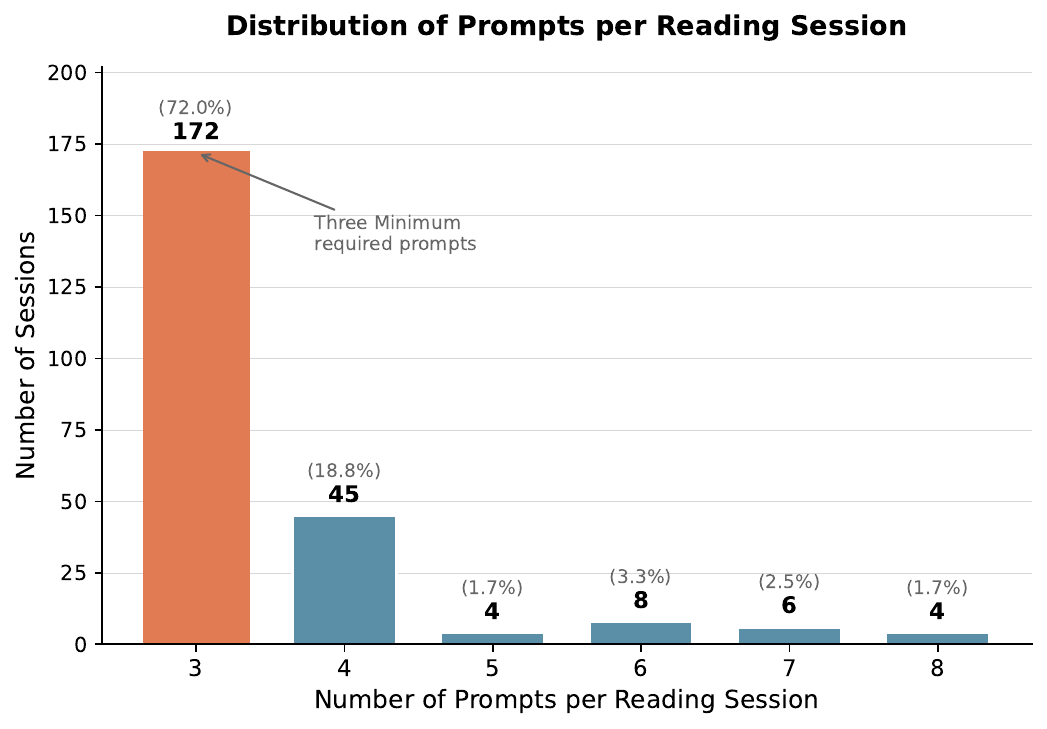}
    \caption{Distribution of prompts per reading session ($N$ = 239 sessions). The assignment required a minimum of three conversational turns per session. A large majority of sessions (72.0\%) contained exactly three prompts, while 28.0\% exceeded this minimum. The orange bar indicates the minimum requirement.}
    \label{fig:prompts_per_session}
\end{figure}

On average, each reading session contained 3.51 conversational turns ($SD$ = 1.06, range: 3--8). Notably, 72.0\% of sessions ($n$ = 172) contained exactly three prompts, which was the minimum required by the assignment. Only 28.0\% of sessions ($n$ = 67) exceeded this threshold (see Figure~\ref{fig:prompts_per_session}). This pattern suggests that most students completed the required interactions without pursuing extended dialogue concerning the reading material.
   
\subsection{Comprehension-Focused Prompts Dominated Prompt Requests }
\label{results: codes_and_themes}

\begin{figure}[htbp]
    \centering
    \includegraphics[width=0.70\textwidth]{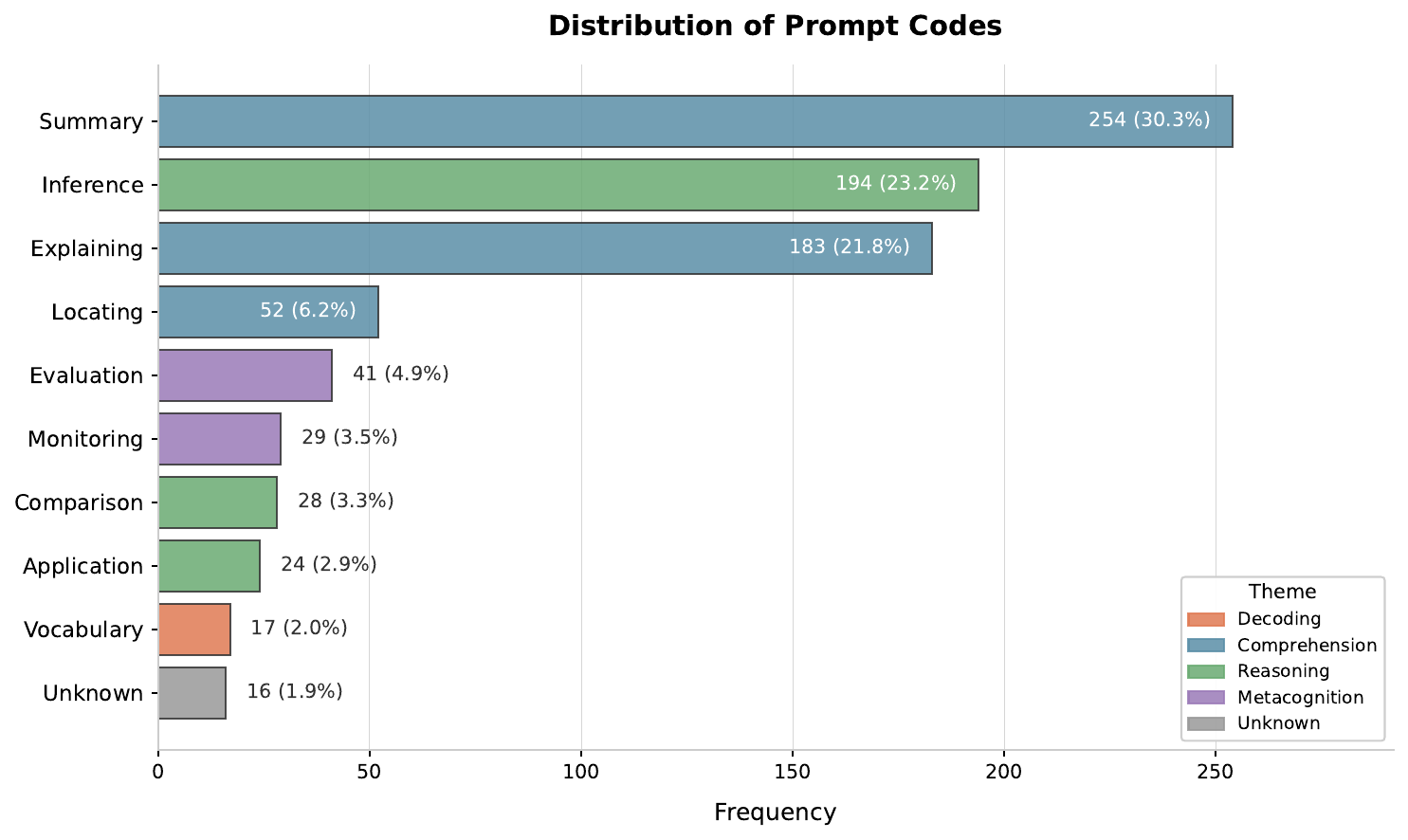}
\caption{Distribution of prompt codes across 838 prompts. Summary (30.3\%), Inference (23.2\%), and Explanation (21.9\%) were the most frequent codes, together accounting for 75.4\% of all prompts. Colors indicate higher-level cognitive themes.}
    \label{fig:code_distribution}
\end{figure}

We coded all 838 prompts according to eleven codes. After excluding the ``unknown'' prompts ($n$ = 16), we then aggregated these 822 codes into four cognitive themes: Decoding, Comprehension, Reasoning, and Metacognition.

\subsubsection{Code-Level Distribution}

At the code level, computed over all 838 prompts including 16 classified as \texttt{UNKNOWN}, three categories dominated: Summary (30.3\%), Inference (23.2\%), and Explanation (21.8\%). Together, these accounted for over three-quarters of all prompts (75.3\%). Metacognitive codes were less frequent, with Evaluation ($n$ = 41, 4.9\%) and Monitoring ($n$ = 29, 3.5\%) together comprising 8.4\% of prompts. Vocabulary requests were the least common ($n$ = 17, 2.0\%). Figure~\ref{fig:code_distribution} displays the full distribution.
\begin{figure}[htbp]
    \centering
    \includegraphics[width=0.70\textwidth]{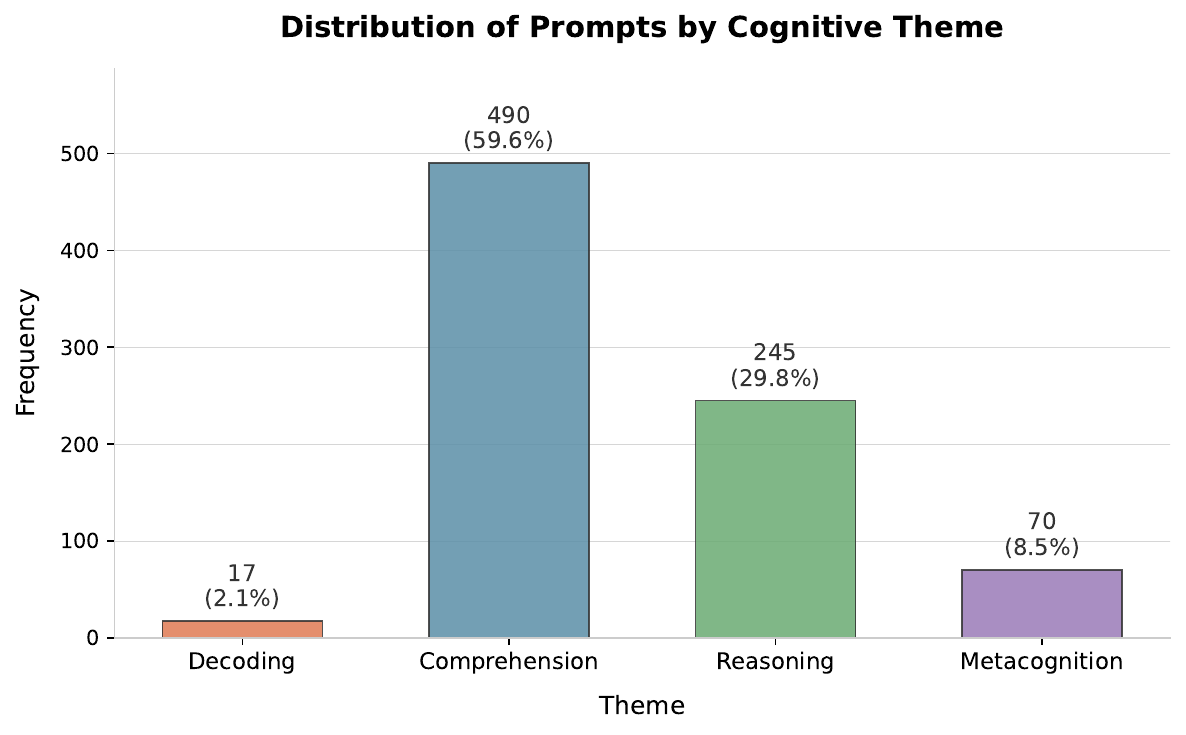}
    \caption{Distribution of prompts by cognitive theme ($n$ = 822 prompts). Comprehension-focused prompts were most prevalent (59.6\%), followed by Reasoning (29.8\%), Metacognition (8.5\%), and Decoding (2.1\%). Themes are ordered from lower-order (left) to higher-order (right) cognitive engagement.}
    \label{fig:theme_distribution}
\end{figure}

\subsubsection{Theme-Level Distribution}

Aggregating codes into themes revealed that Comprehension prompts dominated student interactions, accounting for 59.6\% of all coded prompts ($n$ = 490). Reasoning prompts were the second most common at 29.8\% ($n$ = 245), followed by Metacognition at 8.5\% ($n$ = 70) and Decoding at 2.1\% ($n$ = 17). Figure~\ref{fig:theme_distribution} shows this distribution.
 \subsection{Limited Investment in Prompt Quality and Extended Interaction}
\label{results:limited_investment}
\begin{figure}[htbp]
    \centering
    \includegraphics[width=0.70\textwidth]{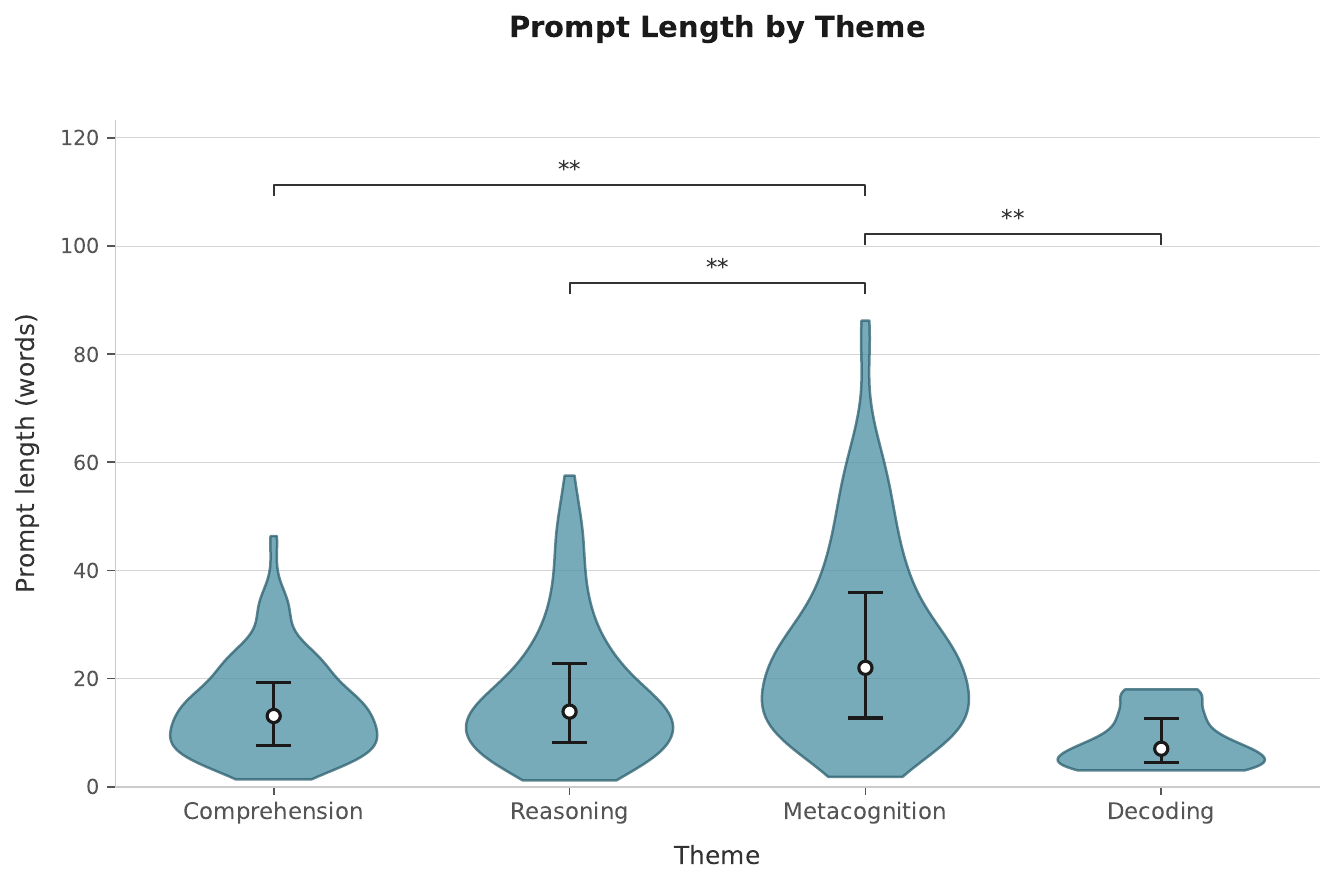}
    \caption{Distribution of prompt length (in words) across four cognitive themes. Data were trimmed at the 99th percentile to exclude outliers. Significance brackets indicate pairwise comparisons from a linear mixed-effects model. Metacognitive prompts were significantly longer than Comprehension, Reasoning, and Decoding prompts. **p < .01. }
    \label{fig:prompt_length_model}
\end{figure}
Beyond the prevalence of comprehension-focused prompts, students showed limited investment in the quality and depth of their AI interactions. Three complementary findings support this characterization: prompt brevity, minimal use of taught prompting engineering strategies, and low responsiveness to AI-generated prompt suggestions.

\subsubsection{Most Prompts Were Brief, Though Metacognitive Prompts Required Longer Elaboration}

The average prompt contained 16.1 words ($SD$ = 11.9, $Median$ = 13.0), with most prompts being concise requests. A linear mixed-effects model with cognitive theme as a fixed effect and participant as a random intercept revealed that metacognitive prompts were significantly longer than prompts in other themes. Using Metacognition as the reference category ($\beta$ = 20.03, $SE$ = 2.28, $p$ < .001), Comprehension prompts were 4.4 words shorter ($\beta$ = --4.37, $SE$ = 1.26, $p$ < .01), Reasoning prompts were 4.1 words shorter ($\beta$ = --4.13, $SE$ = 1.32, $p$ < .01), and Decoding prompts were 7.2 words shorter ($\beta$ = --7.19, $SE$ = 2.61, $p$ < .01). See Figure~\ref{fig:prompt_length_model}.

Qualitative review revealed that metacognitive prompts often included substantial context. For example, one of a participant's metacognitive prompt is ``\textit{The author argues that AI should be treated like a person because it behaves like one in certain ways, despite not being sentient. Critique this argument. Does this analogy hold, or does it risk misleading users about the true nature of AI?}'' (42 words, P6). This pattern indicates that metacognitive engagement requires greater effort to articulate one's thinking.

\subsubsection{Students Rarely Applied Taught Prompting Strategies}
\label{results:not_apply_prompt_engineer}
During the first week, the instructor introduced the CO-STAR prompt engineering framework (Context, Objective, Style, Tone, Audience, Response format~\cite{costar, teo_2023}) and emphasized effective prompting throughout the course. Despite this instruction and students' own recognition that good prompts yielded better responses (see Section~\ref{results:prompt_critical}), they rarely applied these techniques in practice.

Of 822 codable prompts, only 35 (4.3\%) incorporated any prompt engineering element beyond a basic objective (e.g., ``summarize this paper''). Using the CO-STAR  \cite{costar} prompting engineering framework as our coding lens, we found that Audience specifications were the most common prompt engineering element (2.2\%), such as ``\textit{explain it in a way that a freshman in university would understand}.'' Response format requests were also relatively common (1.8\%), such as ``\textit{provide this in a more digestible format}.'' Style (1.2\%) and Tone (0.7\%) were rare. Nearly half of participants (7 of 15, 47\%) never used any prompt engineering techniques across the entire study; among those who did, usage ranged from 2\% to 14\% of their prompts.

\subsubsection{Students Rarely Followed AI-Generated Suggestions}
\label{results:rarely_followup}
Students also showed limited responsiveness to AI scaffolding. Across 239 sessions, 64 sessions (134 instances) included AI-generated invitations for follow-up at the end of responses. Most invitations ($n$ = 84, 62.7\%) were generic (e.g., ``\textit{Would you like further detail on any of these aspects?}''). Only 50 instances (37.3\%) contained specific suggestions, such as ``\textit{Would you like to discuss how to make AI-assisted work more original and personalized?}''

Even when AI provided specific suggestions, students followed only 10 of 50 (20\%). This behavior was concentrated in two participants: P10 followed 7 of 11 suggestions, and P9 followed 3 of 12. The remaining six participants who received specific suggestions followed none.

Together, these findings characterize a task-completion orientation toward AI interaction. Students wrote brief prompts, rarely applied prompting strategies they had learned, and seldom pursued AI-suggested follow-ups. This pattern persisted despite explicit instruction and despite students' stated awareness that more thoughtful prompting yielded better results.

\subsection{Cognitive Progression Within Sessions}

\begin{table}[htbp]
\centering
\caption{Theme distribution by prompt position (\%).}
\label{tab:theme_by_position}
\begin{tabular}{lrrrr}
\toprule
Position & Decoding & Comprehension & Reasoning & Metacognition  \\
\midrule
1st & 0.8 & 90.3 & 7.2 & 1.7 \\
2nd & 2.6 & 53.6 & 32.8 & 11.1  \\
3rd & 3.0 & 36.2 & 49.4 & 11.5  \\
4th+ & 1.7 & 56.9 & 30.2 & 11.2 \\
\bottomrule
\end{tabular}
\end{table}

\begin{figure}[htbp]
    \centering
    \includegraphics[width=0.85\textwidth]{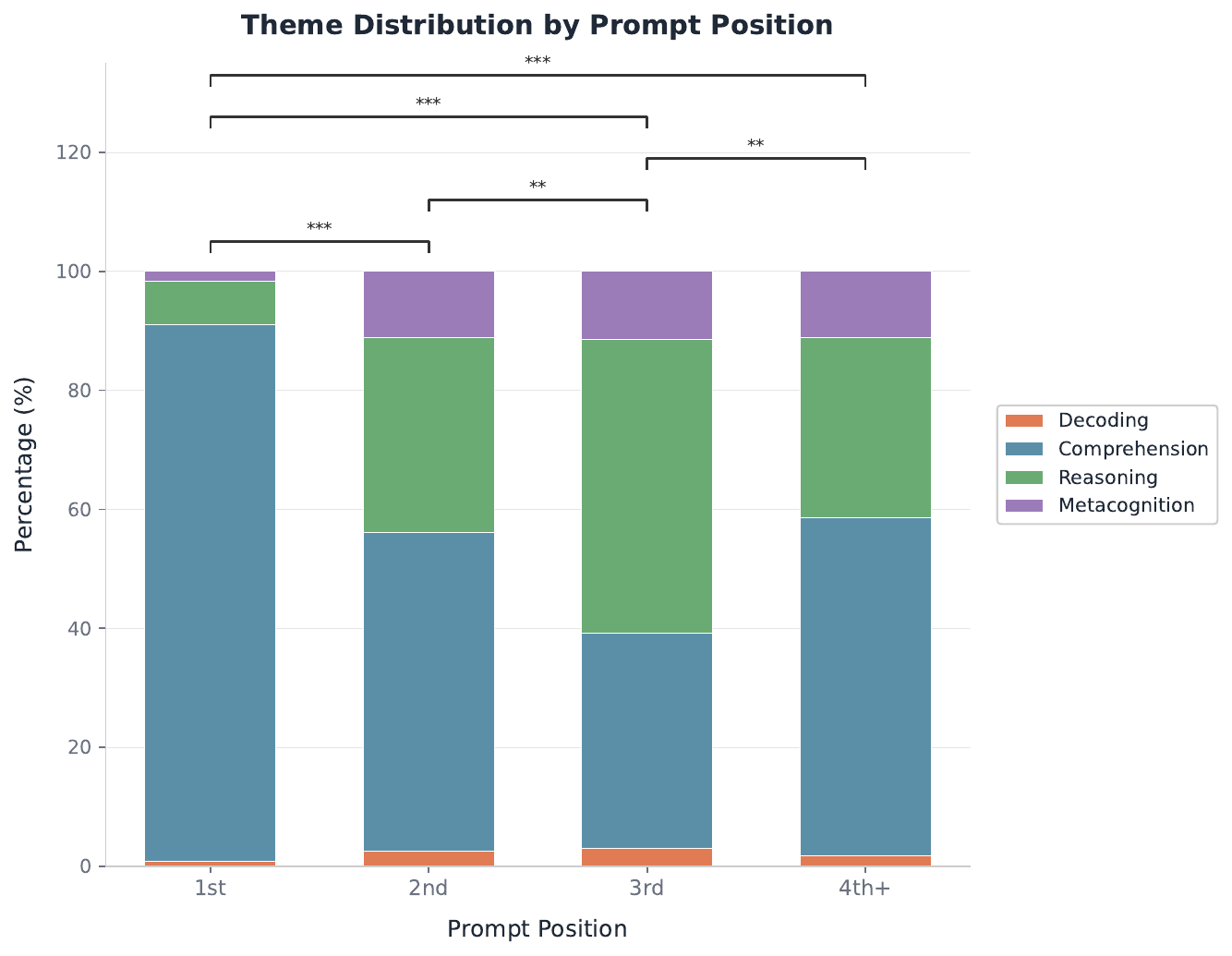}
    \caption{Theme distribution by prompt position within reading sessions. The first prompt was predominantly comprehension-focused, with reasoning prompts increasing substantially by third prompt. Significance brackets indicate pairwise chi-square comparisons with Bonferroni correction; ** $p$ < .01, *** $p$ < .001.}
    \label{fig:theme_by_position}
\end{figure}

\begin{figure}[htbp]
    \centering
    \includegraphics[width=1\textwidth]{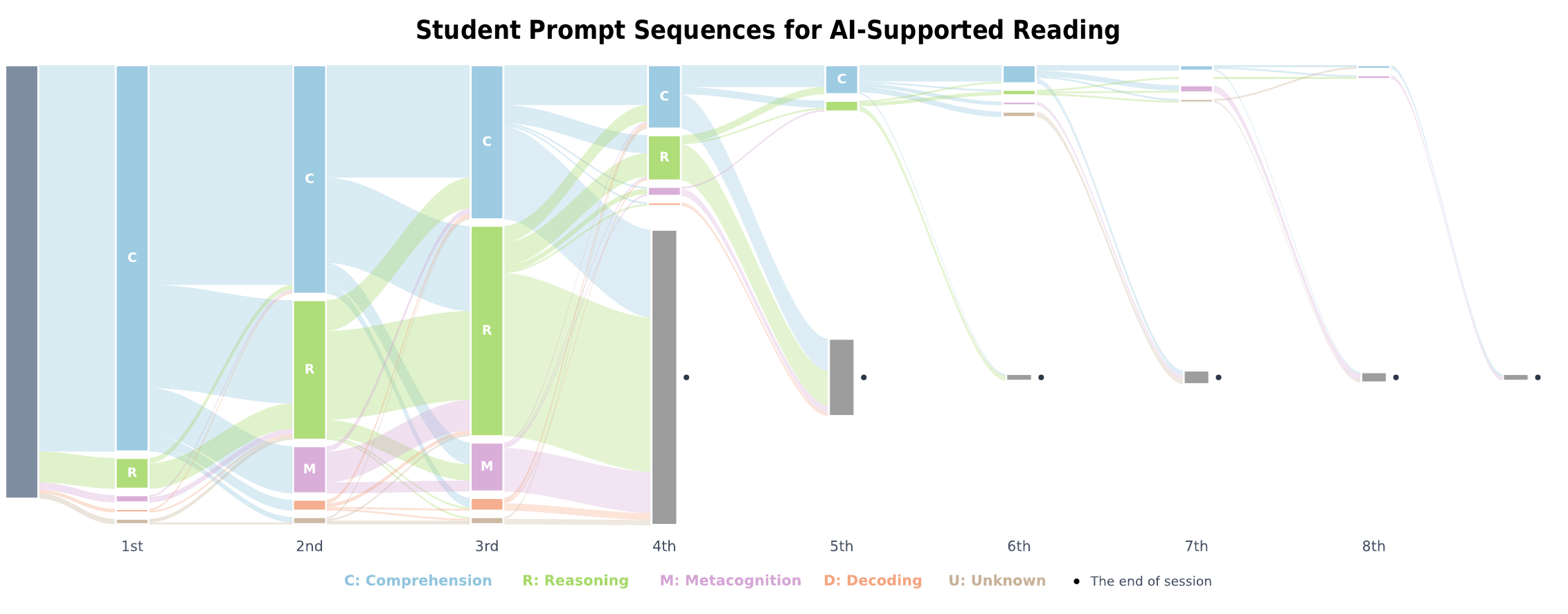}
    \caption{Sankey diagram showing sequential flows of themes across prompt positions. Node heights reflect the number of prompts at each position.}
    \label{fig:alluvial}
\end{figure}

To examine students' cognitive engagement within reading sessions, we analyzed the distribution of themes across prompt positions  (see Table \ref{tab:theme_by_position}). The pattern was illustrated in Figure \ref{fig:alluvial}, a Sankey diagram showing the sequential flow of cognitive themes across prompt positions.  The 1st prompt was overwhelmingly comprehension-focused (90.3\%), with students typically initiating AI interactions by seeking summaries or explanations of the reading material. By the 2nd position, the distribution shifted substantially: Comprehension declined to 53.6\% while Reasoning increased to 32.8\%. This trend continued through the 3rd position, where Reasoning prompts (49.4\%) exceeded Comprehension prompts (36.2\%) for the first time. Metacognition remained relatively stable across positions 2nd–4th+ (approximately 11\%). 

A chi-square test of independence revealed a significant association between prompt position and cognitive theme ( $\chi^2(9)$ = 152.78, $p$ < .001, Cramér's $V$ = .25). See Figure \ref{fig:theme_by_position} for a visual distribution. Post-hoc pairwise comparisons with Bonferroni correction showed that the 1st prompt's cognitive theme distribution differed significantly from all subsequent positions (all $p$ < .001). The 2nd and 3rd prompts also differed significantly in terms of theme distribution ($p$ < .01), as did the third and fourth+ positions ($p$ < .01). Notably, the 2nd and 4th+ positions did not differ significantly ($p$ = .907), suggesting a regression toward comprehension engagement pattern in extended sessions. A qualitative review of the 4th+ prompts revealed that students often concluded their sessions by requesting key takeaways or summaries (for example, one final prompt by a participant was ``\textit{Can you summarize the paper again but create less key points?}'' (P9)), which were coded as Summary belonging to theme Comprehension.

\subsection{Consistent Engagement Patterns Across Eight Weeks}
\label{results: across_weeks}
\begin{figure}[htbp]
    \centering
    \includegraphics[width=\textwidth]{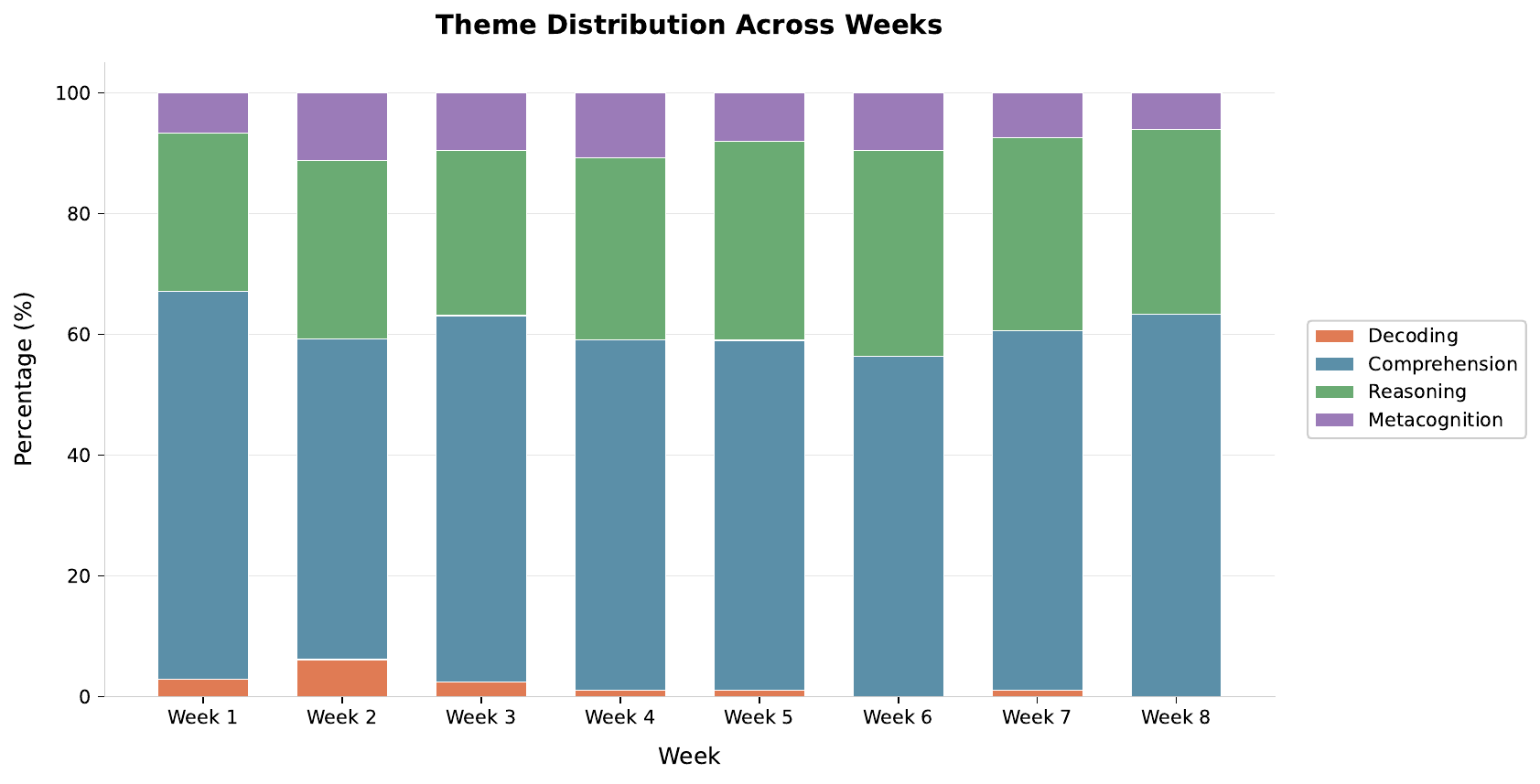}
    \caption{Distribution of cognitive themes across the 8-week study period. Theme proportions remained stable over time, indicating that students' patterns of cognitive engagement with the AI remained stable throughout the study..}
    \label{fig:theme_by_week}
\end{figure}

To examine temporal changes in cognitive engagement, we first tested whether theme distribution varied significantly across the 8-week study period after excluding uncodable prompts ($n$ = 822). A chi-square test of independence revealed no significant association between week and theme distribution ( $\chi^2(21)$ = 19.81, $p$ = .533), suggesting engagement patterns were established early and persisted (see Figure \ref{fig:theme_by_week}). 
To test whether this stability also held at the individual level, we conducted split-half correlations comparing each participant's theme percentages from weeks 1–4 (early) with weeks 5–8 (late). Among participants in both periods, high-frequent theme Comprehension showed moderate-to-strong rank-order stability ($r$ = .69, $p$ < .01), as did Reasoning ($r$ = .63,  $p$ < .05). Students who began the course relying heavily on comprehension support continued that pattern; those who engaged more frequently in reasoning maintained that approach. Metacognition ($r$ = .35, $p$ = .221) and Decoding ($r$ = .17, $p$ = .569) did not show significant stability, suggesting these lower-frequency themes were more context-dependent. 

This persistence is noteworthy given that the course focuses on AI. Throughout the course, students engaged with readings about AI's cognitive influence, risks of overdependence, and the importance of critical engagement with AI tools. Yet their interaction patterns showed no evidence of change. The finding suggests that initial AI interaction habits, once established, may be resistant to change even when course content explicitly addresses effective AI use, raising questions about whether exposure to course material and instruction by the teacher without structured evaluation policy is sufficient to shift students' cognitive engagement in reading tasks with AI.

\subsection{Individual Difference in AI Engagement}
\label{results:individual_difference}
\begin{figure}[htbp]
    \centering
    \includegraphics[width=\textwidth]{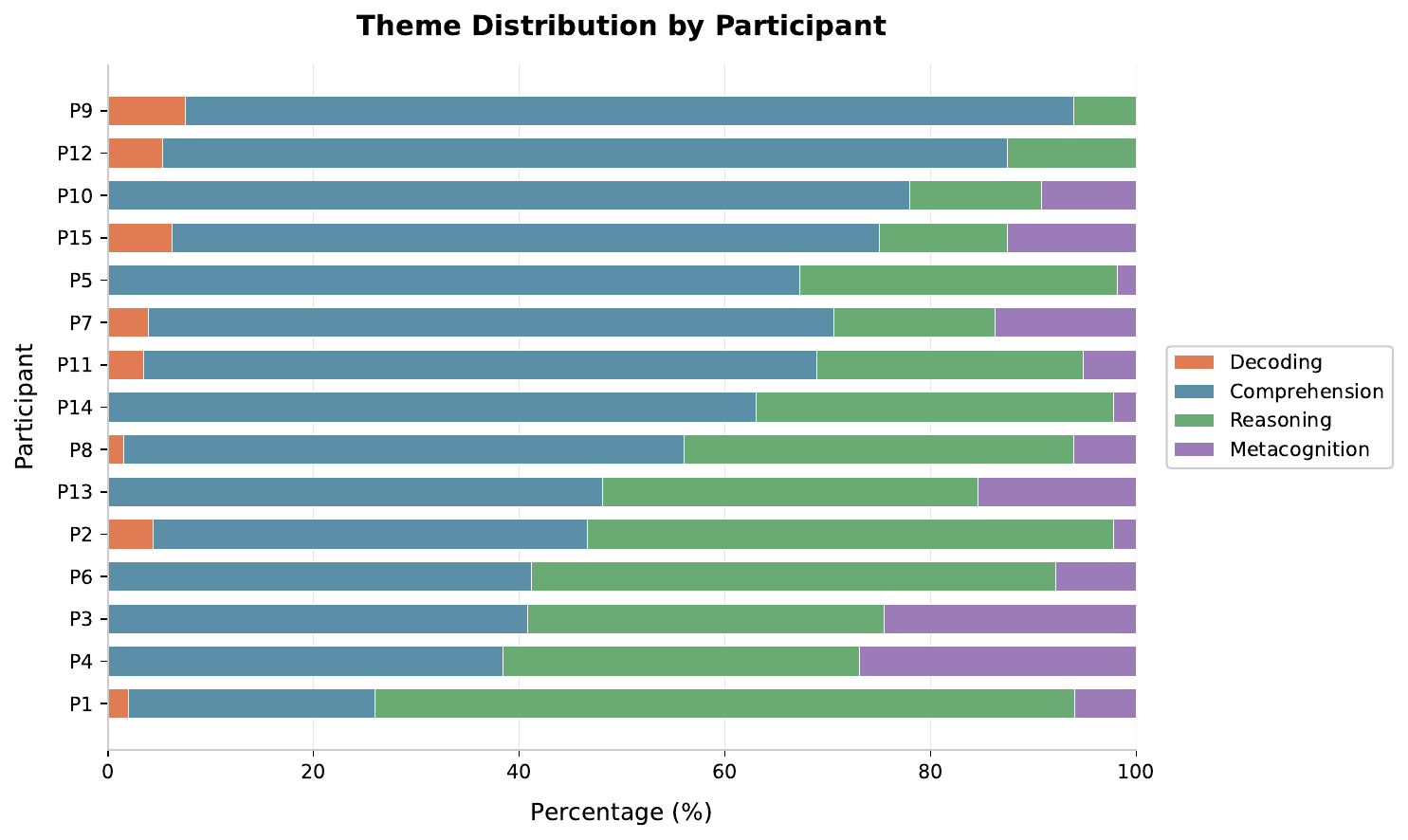}
    \caption{Theme Proportions by Participant. Stacked horizontal bars show the percentage of prompts in each cognitive theme for each participant. Students at the top of the figure engaged primarily in decoding and comprehension, while those at the bottom relied predominantly on reasoning-focused prompts.}
    \label{fig:participants_variance}
\end{figure}

Substantial between-participant variability characterized students' cognitive engagement patterns. Across the 15 participants, Comprehension accounted for an average of 57.8\% of prompts, but individual usage ranged widely from 24.0\% to 86.4\% ($SD$ = 18.2). Similarly, Reasoning averaged 31.0\% but ranged from 6.1\% to 68.0\% ($SD$ = 17.3). Metacognition ($M$ = 8.9\%, $SD$ = 8.4, $range$: 0.0–26.9\%) and Decoding ($M$ = 2.3\%, $SD$ = 2.7, $range$: 0.0–7.6\%) showed lower overall usage but also showed high variability. The results show that students had distinct AI engagement patterns: some relied heavily on comprehension support while others engaged more frequently in reasoning (See Figure \ref{fig:participants_variance}).

To quantify the extent to which cognitive engagement patterns reflected individual differences rather than week-to-week variation, we computed intraclass correlation coefficients (ICC) for each theme using participants' weekly theme percentages. ICCs for Comprehension (.51) and Reasoning (.51) indicated that approximately half of the variance in these themes was attributable to between-participant differences. This suggests that students have distinct AI interaction styles, for example, some consistently relied on comprehension support while others engaged more frequently in reasoning, and these patterns were maintained across the 8-week study period. Metacognition showed an ICC (.31), indicating that about 31\% of variance reflected stable individual differences. Decoding had the lowest ICC (.20). Together, these findings demonstrate that individual differences in AI chatbot usage are not random fluctuation but reflect consistent distinct cognitive engagement profiles that persisted throughout the study.

Prompt length also varied substantially across participants by word count. The average prompt length was 16.1 words ($SD$ = 7.9), but participant means ranged from 8.1 to 40.5 words. Some students consistently wrote brief prompts (e.g., P7: $M$ = 8.1 words; P2: $M$ = 9.3 words), while others provided more elaborate requests (e.g., P3: $M$ = 40.5 words; P6: $M$ = 24.2 words). The ICC for word count was .40, indicating that 40\% of variance in prompt length was attributable to between-participant differences, suggesting that students developed characteristic prompting styles that persisted across sessions.

\section{Understanding Student Perspectives on AI-Supported Reading}

We conducted interviews with five participants and analyzed reflections of AI usage from 15 participants' reading sessions to understand RQ3 (Why do students engage with AI in the ways they do?). To differentiate these two data sets, we employed a labelling system. Reflection of AI usage data received a ``-s'' suffix added to the participant ID (e.g., ``P14s''), whereas for the interview, we used the participant ID without any suffix (e.g., ``P14'').

\subsection{Efficiency as the Primary Driver}

Students consistently identified efficiency as the primary value of AI-supported reading. P9s captured this directly: ``\textit{it again amazes me how these tedious tasks could be completed within 5 minutes thanks to the help of AI.}" Another explained their preference simply: ``\textit{I would use AI more because it's faster, sufficient and I can get answers most of the time it's reliable}" (P3). P7s noted that AI ``\textit{definitely increased my efficiency in covering the topics mentioned.}"

Some students were aware of the distinction between speed and understanding, as P13 distinguished efficiency from comprehension improvement by saying, ``\textit{I wouldn't say it positively helped my reading comprehension...[it] just make my reading more efficient.}" P13 subsequently explained that skimming through the reading before using AI provided ``\textit{preexisting knowledge}" necessary to verify outputs and formulate targeted questions. Another participant echoed this point by describing how this initial engagement transformed the interaction: ``\textit{[skimming] is usually when my deeper questions come into play}" (P1). 
 
However, Other students prioritized efficiency over engagement entirely. P9 admitted: ``\textit{When the paper is really long and I don't really have a lot of time, I kind of ask it to summarize something for me and give me the main points.}" P12 described similar behavior for low-priority coursework: they would ``\textit{put all of that reading in AI, barely try to understand it,}" delegating the cognitive work to the system. Aiming for efficiency, students defaulted to ``\textit{generic questions from what AI is generating,}" resulting in a disconnect from the source material (P13).

The emphasis on efficiency aligns with the quantitative finding that 72\% of sessions contained exactly the required minimum of three prompts. Students optimized for task completion rather than extended exploration, a pattern consistent with their stated priority of saving time.

\subsection{The Intention-Behavior Gap in Effective Prompting}
\label{results:prompt_critical}
Across participants, they emphasized the importance of crafting good prompts. P1 articulated a clear distinction, noting that while ``\textit{surface level prompts just gives me a basic answer,}" crafting deeper prompts was essential to ``\textit{evaluate why that answer was given and how it can apply to different scenarios}." They recognized crafting intentional prompts was essential to critically engagement of the reading, as one participant noted, ``\textit{formulating my last prompt took some critical thinking}'' (P12s), showing prompting itself as a cognitive activity rather than mere information retrieval. Similarly, another participant emphasized ``\textit{Some of the most effective prompts I used were the ones that helped me critically assess the validity of the article}'' (P4s). 

Additionally, students were aware of the importance asking multi-turn follow-up questions. One participant said, ``\textit{prompting AI with multiple [follow-up] questions allowed me to critically analyze assumptions and biases}'' (P4s). And P12 summarized clearly: 
\begin{quote}
    ``\textit{The way I prompted and seeing the outputs that I would get...And it felt like there was never a one single response you could prompt AI and get everything out of. That's what I truly realized when I was doing [correctly of] all of this [prompting], that you have to constantly prompt AI questions about the reading to fully understand it, because with that one summarizing prompt, you're not going to get everything out of it}" (P12).
\end{quote}

Students also identified specific prompting strategies that deepened their understanding and engagement. 1) Requesting concrete examples helped ground abstract concepts and connect to students' own lives: ``\textit{I found it helpful to ask for real life examples}" (P2s), and ``\textit{[AI] gives me some real life examples that I can relate with it}" (P12). 2) Asking implication and comparison prompts proved particularly effective ``\textit{the prompts that worked best for me were the ones that asked AI to compare}'' (P4s). 3) Some discovered novel approaches, such as asking ``\textit{What's something I might find surprising from the article?}" a prompt the student thought elicited especially engaging responses (P15s). Taken together, these showed students are cognizant ``\textit{the better your questions are, the more helpful AI can be}'' (P12s).

Despite this recognition that effective prompting required effort and yielded better results, quantitative analysis revealed that only 4.3\% of prompts incorporated prompt engineering elements beyond basic objectives and 72\% of sessions contained only three prompts (see Section \ref{results:limited_investment}). This intention-behavior gap between stated understanding and actual practice suggests that while students intellectually grasped the value of crafting good prompting, they were not able to consistently apply it. 
       
\subsection{Strategic Triage in AI-Supported Reading}
       
Students described strategic \textit{triage} of their reading engagement, operating at two levels: across coursework and within individual readings. AI served as the primary mechanism for both.

\subsubsection{Triage Across Coursework}

The depth of AI engagement depended on personal interest and academic pressures. For low-priority classes, AI functioned as a bypass mechanism. P12 admitted that for ``\textit{don't care}'' classes, the goal was meeting grading criteria with minimal cognitive load. In contrast, genuine interest prompted deeper engagement. P1 noted that interesting topics led them to ``\textit{take a little bit more time}'' reading manually before asking deeper questions about the author's intent or societal impact.

Academic pressures further shaped this triage. Deadlines forced utilitarian shifts away from deep reading. Even students who valued learning admitted reverting to summarization under time constraints: ``\textit{I could have used AI maybe a little bit less... But at that time, I didn't have enough time}'' (P12). External pressures often obliged passive usage regardless of students' stated preferences.

\subsubsection{Triage Within Readings}
Beyond deciding which readings deserved attention, students used AI to triage \textit{within} readings by identifying which sections warranted deeper engagement. AI-generated summaries served as filtering mechanisms, allowing students to scan content and selectively pursue points of interest:
\begin{quote}
``\textit{AI helped me get a basic understanding of this article by breaking it up into 9 main points. This helped me quickly skim through and identify which points I would need to read further upon. For example, I was more interested in seeing how the author's position and argument which were covered in points 4--6. I was able to get further clarification on these points by asking ChatGPT to go into more detail}'' (P11s).
\end{quote}

Two prompt types functioned as triage mechanisms within this new phenomenon. First, \textit{locating prompts} allowed students to probe whether the text addressed specific topics without reading it themselves (see code \texttt{LOC} in Table~\ref{tab:prompt_codes}). These prompts typically began: ``\textit{Does the article mention/include/talk about...}'' Students used them to connect prior assumptions or interests to reading content, filtering for relevance before investing effort. Second, \textit{vocabulary prompts} addressed unfamiliar terms encountered in AI's summaries rather than the original text (see code \texttt{VOCAB} in Table~\ref{tab:prompt_codes}). P13s described: ``\textit{I find learning from the summary AI make for me and continue asking questions that related to its response by using quotation marks for the exact words.}''

These patterns indicate a shift from traditional reading, in which direct engagement with the text is required to identify relevant content and unfamiliar terms. Students effectively read \textit{through} AI rather than \textit{with} it: using AI output as the primary material to process and triage, with the original text serving as a secondary resource consulted selectively. 
 
\subsection{Individual Differences in Engagement}
Our quantitative analysis showed that participants varied substantially in how they engaged with AI, ranging from comprehension-heavy to reasoning-heavy profiles (Section~\ref{results:individual_difference}, Figure~\ref{fig:participants_variance}). Although our interview sample is small, the interviewed participants' accounts broadly tracked their profiles. Comprehension-heavy participants framed AI around efficiency: P9 explained that ``\textit{if you have a summary in front of you, that's kind of mostly what you need in order to succeed in the class,}'' and P12 described low-interest courses as cases where ``\textit{I will tend to use a lot of AI... because it's quick work.}" Other participants described the opposite orientation, doing the thinking first and using AI to check or extend it. P1, the most reasoning-oriented participant, noted that ``\textit{if I really care about the subject, I'll do my best to not use AI.}'' These accounts suggest that the individual differences may reflect stable orientations, shaped by how each student balanced efficiency, interest, and self-regulation when deciding how deeply to engage with AI-supported reading.

\subsection{Tensions in AI-Supported Reading }

Students articulated both benefits and concerns about AI-supported reading. Notably, many concerns directly mirrored perceived benefits, revealing inherent tensions in AI-supported reading.

\subsubsection{Convenience vs.\ Shallow Processing, Procrastination, Skill Erosion, and Overdependence}

Students valued AI's ability to quickly simplify complex material and reduce cognitive load. P9s found scientific papers with ``\textit{many technical terms}'' challenging, noting: ``\textit{Having AI explain it to me in simple terms made things a lot easier.}'' For P12, a student with dyslexia, AI addressed significant comprehension barriers: ``\textit{I have dyslexia and usually it's a lot harder for me to understand and read information... ChatGPT would break down each thing and actually give me examples of the definitions.}'' Students also appreciated AI's flexible output formats, such as bullet points or specific word counts. P5s noted: ``\textit{I love how they use bullet points to explain. By using bullet points, users will clearly understand what the main points are.}''

However, this convenience generated multiple concerns. First, students recognized that efficiency came at the cost of deep processing. P13 observed that relying on summaries ``\textit{takes away from how much you get out of that summary,}'' removing the struggle of ``\textit{processing your mind for yourself.}'' P9 noted: ``\textit{Since the summarization happens quicker... it's maybe less engagement.}''

Second, convenience paradoxically fostered procrastination. P12 explained: ``\textit{It makes me procrastinate more and make me lose that motivation a bit more rather than trying to quickly get it done.}'' Knowing AI could summarize readings in seconds allowed students to delay work until the last minute.

Third, students expressed anxiety about skill erosion. P12 reported noticing personal decline: ``\textit{I started using AI for a good two years and relying on it for reading completely. And I started to lose... my ability to analyze stuff in an analytical way.}'' This concern extended to overdependence more broadly. P13 warned: ``\textit{You can rely on it so much that it becomes almost a fallback for your learning.}'' The same convenience that reduced immediate burden raised concerns about long-term cognitive costs.

\subsubsection{Personal Relevance vs.\ Not Forming Own Opinions}
\label{results:personal_relevance}
Students valued AI's ability to personalize content and make readings relevant to their lives. Unlike static textbooks, AI allowed students to explore personally meaningful questions. One participant found asking ``\textit{What suggestions would you give to the current college students regarding the issues you mentioned?}'' (P13) helpful for situating readings in their own context. Students could request explanations tailored to their comprehension level and connect abstract concepts to their future careers or personal experiences.

Yet this personalization came with a cost: students found AI could hinder having opinions by the readers themselves. P8s noted: ``\textit{It helps to learn what the article is about, but not enough to form my own opinion about it. I had to read some of the parts to actually form an opinion on this.}'' P1 echoed: ``\textit{It does [affect] your ability to think of ideas on the spot.}'' When AI pre-processed and framed content, students had fewer opportunities to develop their own interpretations.

\subsubsection{Extended Knowledge vs.\ Missing Information}

Students appreciated AI's ability to extend beyond the assigned text. P11s valued when AI ``\textit{went beyond what was in the article and did an online search/data synthesis.}'' P13s echoed: ``\textit{AI provided me with external sources to help me answer the question... This is an efficient use of recourse and integrates various information to generate a comprehensive answer.}'' This capability allowed students to situate readings within broader contexts and access supplementary information efficiently.

However, students worried about what AI omitted. P10s noted: ``\textit{While AI is good at shortening long material, it tends to lose important information in the process.}'' The same capability that extended knowledge also risked narrowing it---AI's selective summarization meant students might miss content that fell outside the system's filtering.

Together, these tensions reveal a consistent pattern: the features students valued most---convenience, personalization, and extended knowledge---simultaneously created risks of shallow processing, diminished independent thinking, and incomplete understanding. Students were aware of these trade-offs, yet behavioral patterns suggest that immediate benefits typically outweighed long-term concerns in practice.

\section{Discussion}

\subsection{The Cognitive Shape of AI-Supported Reading: Outsourcing Comprehension and Truncating Progression}

Our findings show that, rather than using AI as a springboard for deeper analysis, students primarily used it to outsource comprehension itself. Comprehension prompts dominated the dataset (59.6\%), with Summary requests alone accounting for 30.3\% of all prompts. Reasoning was less frequent (29.8\%), Metacognition rare (8.5\%), and Decoding nearly absent (2.1\%). Contrary to the expectation that AI would serve as a Socratic tutor, the dominance of ``Summary" and ``Explanation" codes suggests students largely used AI to distill complex information into a digestible format.

This pattern is concerning in light of text comprehension theory. Kintsch's construction-integration model distinguishes between the \textit{textbase}, a representation of the text's explicit content, and the \textit{situation model}, a deeper representation that integrates text meaning with prior knowledge~\cite{kintsch1988role}. Building a situation model requires active inference, elaboration, and connection-making. When students delegate summarization and explanation to AI, they may be receiving a pre-constructed textbase without performing the integrative cognitive work that supports durable understanding. In short, they bypass the productive struggle of meaning-making.

Despite this tendency, our within-session analysis reveals that students possess a natural inclination toward higher-order engagement. As shown in Figure~\ref{fig:alluvial}, a clear cognitive progression emerged: students began with comprehension-focused prompts (90.3\% at the first position) but shifted toward reasoning as sessions continued. By the third prompt, Reasoning (49.4\%) exceeded Comprehension (36.2\%) for the first time. This pattern suggests that when interaction continues, students naturally move from building foundational understanding toward interrogating and reasoning.

However, this progression was routinely cut short. Because 72\% of sessions contained exactly three prompts, the assignment minimum, most students exited precisely as they reached the threshold of higher-order engagement. The minimum requirement, intended as a floor, functioned as a ceiling. Students completed the cognitive ramp-up but did not sustain it.

We argue this truncation reflects not a lack of student capability but a limitation of general-purpose AI systems. Current chatbots function as passive respondents rather than active tutors. Unlike a human instructor who might push back after a summary by asking, ``What do \textit{you} think about that?,'' AI provides a comprehensive answer and halts. It does not nudge the user to elaborate, challenge the user's assumptions, or sustain the dialogue. By failing to scaffold the user to continue engaging with the reading material, the system implicitly signals that the learning task has been completed. The user's cognitive progression is thus truncated by a system architecture that optimizes for immediate task completion over sustained inquiry.

These findings suggest opportunities for structured scaffolding of cognitive progression. AI reading systems could guide students through phases (e.g., first building comprehension, then prompting reasoning, and finally encouraging metacognitive reflection) rather than responding passively to whatever students request. Additionally, teacher-in-the-loop systems could enable instructors to set parameters for specific readings: minimum interaction duration, required prompt types, or learning objectives that the system reinforces through suggested questions. Monitoring dashboards that show class-level engagement patterns would enable instructors to adjust scaffolding as a course progresses, identifying when students consistently exit before deeper engagement. The goal is not to mandate more prompts but to structure interactions so that cognitive progression is supported rather than truncated.

\subsection{Efficiency and Epistemic Risk: Reading Through AI Rather Than With It}
\label{dis:reading_through_ai}

Efficiency emerged as the dominant driver of students' AI use. Students stopped at the assignment minimum (72\% of sessions contained exactly three prompts), rarely applied taught prompting strategies (4.3\%), and ignored 80\% of AI-generated follow-up suggestions. These patterns align with research on strategic reading in higher education, which documents that students routinely calibrate effort based on time constraints, grade implications, and perceived payoff~\cite{hollander2022importance}. AI appears to amplify these tendencies by reducing the marginal cost of satisficing: when a single prompt yields an adequate summary, deeper engagement becomes an optional luxury that students feel they do not have time for.

Our findings reveal that efficiency-driven behavior does more than truncate engagement. It fundamentally restructures how students relate to texts. Two prompt patterns illustrate this shift. First, students frequently used \textit{locating prompts} (\texttt{LOC}), asking questions like ``Does the article mention X?'' or ``Does the author discuss Y?'' without having read the text themselves. Second, students asked \textit{vocabulary prompts} (\texttt{VOCAB}) about unfamiliar terms that appeared in AI-generated summaries, not in the original text. In both cases, students were processing AI's output as if it, and not the assigned reading, was the primary source to examine. The text becomes a background resource consulted selectively, if at all, while AI output serves as the foreground object of comprehension.

This shift carries epistemic risks. By using locating prompts to probe whether the text addresses their pre-existing interests, students create a self-reinforcing filter. They see only what they already thought to ask about, missing the serendipitous discovery of new ideas that occurs in linear reading. A student asking ``\textit{Does this article discuss ethical implications?}'' will learn whether ethical implications are mentioned but will not encounter the unexpected methodological insight or historical context that might have reshaped their thinking. The reading experience becomes confirmatory rather than exploratory.
  This filtering raises a deeper concern: students may not recognize how much they have missed. Students may feel they have read the paper when they have only read their own query-filtered version of it. Worse, this filtered version can compound their confidence, so that what they believe they have learned outruns what they have actually absorbed. The learning sciences term this an \textit{illusion of competence}, in which learners overestimate how well they understand material~\cite{bjork1999f}. A common source of this illusion is \textit{processing fluency}, the subjective ease or familiarity a learner feels when working with material. When information is easy to process, as with an AI-generated summary, learners may tend to treat that ease as a sign of genuine understanding, even when it is not. Prior work shows that learners inflate judgments of their own learning when an answer is visible during study but absent at test~\cite{koriat2005illusions}, and that the fluency of rereading highlighted text can potentially lead learners to mistake that familiarity for learning~\cite{yue2015highlighting}.

AI-supported reading introduces a new source of the illusion. Our findings point to an antecedent that we describe as an \textit{illusion of coverage}. The illusion of competence concerns the self (how well I understand the material), whereas the illusion of coverage concerns the source (belief that the AI's summary represents the reading completely). Students could overestimate how much of the reading the AI's account captured, and three prompt patterns signal why. Students used locating prompts (\texttt{LOC}) to check whether a text addressed their existing interests, asked vocabulary questions (\texttt{VOCAB}) about terms drawn from the AI's summary rather than the original, and rarely monitored their own understanding (metacognitive prompts, \texttt{EVAL} and \texttt{MON}, were only 8.5\% of the total).

Together, these behaviors let students treat the AI's output as a proxy for the reading, so that what the summary omitted went unnoticed. In this way, an illusion of coverage feeds an illusion of competence. This is the cognitive cost of reading \textit{through} AI rather than \textit{with} it: students optimize for targeted retrieval from the AI's account at the expense of the open-ended exploration of the source that characterizes deep reading.

Recent HCI systems try to tackle the problem by pointing toward designs that preserve the connection between AI support and the source text. For example, Paper Plain makes medical research papers more approachable through in-situ term definitions, plain-language section summaries, key questions that guide readers to relevant passages \cite{august2023paper}. Similarly, AI Margin Notes shows that readers prefer LLM support when it appears as document comments linked to manually selected text compared to a conventional chatbot interface separated from the document~\cite{joshi2026designing}. Building on this direction, we suggest several other design interventions could address the risks we observed. First, systems could visualize which concepts a student has explored and proactively surface unexplored areas, prompting students with suggestions like ``You have focused on the methodology. Would you like to explore the theoretical implications or limitations?'' Such nudges could counteract premature narrowing. Second, similar to \cite{august2023paper}, AI responses should be grounded in and linked to specific source passages, encouraging students to cross-reference AI output against the original text. Third, teacher-in-the-loop features could allow instructors to specify key concepts the system should surface regardless of student queries, ensuring that important ideas are encountered even when students do not think to ask about them. Finally, end-of-session summaries that show the breadth of concepts explored and flag what remains unexamined could build metacognitive awareness of coverage, helping students recognize the difference between reading a text and reading their own query-filtered version of it.

\subsection{The Intention-Behavior Gap: Why Knowing Better Does Not Lead to Doing Better}
\label{dis:intention_behavior}

One of the striking findings from this study is the disconnect between students' descriptions of how to learn and their actual behavior. In interviews and reflections, students articulated clear awareness that effective prompting required effort, that multi-turn interactions yielded richer understanding, and that surface-level prompts produced only ``basic answers'' (P1). They recognized that good prompts were very important (P12s). Yet their behavior told a different story. Only 4.3\% of prompts incorporated taught prompting strategies, 72\% of sessions stopped at the minimum requirement, and 80\% of AI-generated follow-up suggestions were ignored. Students knew what good engagement looked like but did not enact it.

This gap persisted despite sustained instruction. Throughout the course, students engaged with readings about AI's cognitive influence, risks of overdependence, and strategies for effective use. The instructor taught prompt engineering techniques during the first week and reinforced them throughout the quarter. Yet we observed no group-level change in engagement patterns across the eight-week study period. Students who began as heavy comprehension users remained so; those who engaged more frequently in reasoning maintained that approach. Exposure to information about effective AI use, even in a course explicitly focused on AI, did not shift behavior. Our findings also show students developed characteristic interaction profiles early and maintained them throughout the study. 

What explains this persistence? We posit three contributing factors. First, external pressures drove strategic triage. Students faced competing deadlines and calibrated their effort based on perceived payoff, a pattern well-documented in research on college reading~\cite{hollander2022importance, lepore2019read}. When a single summary prompt satisfies assignment requirements, deeper engagement becomes discretionary. Second, general-purpose AI systems lack affordances for self-regulated learning. Unlike purpose-built educational tools that scaffold goal-setting, planning, and reflection~\cite{pan2025effects}, chatbots respond to queries without prompting users to evaluate their understanding or extend their thinking. The system architecture optimizes for task completion, not learning. Third, intrinsic motivation varied widely. As an elective course, the class attracted students with different orientations: some enrolled out of genuine curiosity, others primarily for credit. Without personal investment in the material, students approached readings instrumentally regardless of what they knew about effective strategies.

These findings suggest that bridging the intention-behavior gap requires more than instruction. If knowledge alone changed behavior, students would have improved over eight weeks. They did not. Intervention must target the structural and motivational factors that sustain shallow engagement. On the structural side, AI systems could scaffold self-regulation by prompting students to set learning goals before reading and periodically asking them to assess their understanding. Such features would externalize the metacognitive monitoring that students currently neglect. Student-facing dashboards showing prompt types, session duration, and coverage breadth could build awareness of their own patterns over time.

On the motivational side, fostering agency may be key. Research on self-determination theory suggests that learners feel commitment to decisions in proportion to their participation in making them~\cite{knowles1989making}. Offering students choice in reading selection, even limited choice among several options with brief descriptions, could enhance ownership and intrinsic motivation. For required readings where choice is not possible, instructors could make the rationale explicit: why this reading matters, how it connects to course goals, and what students should take away. AI systems could further support relevance by connecting reading content to students' stated interests, career goals, or current concerns, a feature students valued when it occurred (Section~\ref{results:personal_relevance}).

\subsection{Study Context and Transferability}
Our study examined a specific context. The readings were expository texts on AI-related topics, and students received credit for submitting their entries in good faith rather than being tested on comprehension. The course was about AI and might have shaped students' stated beliefs. Each of these factors could shape how students engaged with AI. Below, we discuss where our findings may transfer and how future work can build on them.

Although our setting is specific, the task at its center is not. Preparing for class by reading assigned texts is one of the most common forms of content-area literacy, in which students use reading to learn subject matter across disciplines \cite{shanahan2012disciplinary}. Decades of research show that students do the minimum on such reading, with compliance typically low~\cite{burchfield2000compliance, baier2011college}. The satisficing we observed, where most students stopped at the assignment minimum, fits this long-standing pattern. What AI changes is not the underlying tendency but the ease of acting on it, since a single prompt could fulfil the assignment requirement (e.g. a reading response, a discussion question) with little effort. We therefore expect the reading with AI patterns we surfaced will hold across the many courses that assign reading in this way.
 In addition, these patterns are not unique to our study. A growing body of research shows that AI use can reduce engagement through offloading, the delegation of mental work. Fan et al. found that learners supported by ChatGPT improved short-term performance but showed signs of metacognitive laziness, offloading the self-regulatory work that builds durable understanding~\cite{fan2025beware}, and broader studies link heavier AI use to weaker critical thinking through the same mechanism~\cite{gerlich2025ai}. Our results extend this literature to college reading and show that the similar pattern exists even among students drawn to studying AI, who might have been expected to use the tools more deliberately after being instructed. Left to general-purpose chatbots, it may deepen the existing compliance problem by making disengaged reading frictionless. It is educators and designers' responsibility to design deliberate systems and tasks to motivate and scaffold engagement, helping students read to understand and retain material. 

We acknowledge other contexts may differ. Reading is not a single uniform skill, and the practices that define it vary across disciplines~\cite{shanahan2008teaching, shanahan2012disciplinary}. Readers in history, literature, mathematics, and the sciences engage different kinds of texts in different ways~\cite{shanahan2008teaching}. Where reading centers on direct, close engagement with a primary text, such as a poem or a historical document, reading through an AI summary is less viable and perhaps less tempting, so our patterns may not hold. Also, for technical and mathematical material, students may be pushed toward more deliberate prompting and verification to prepare for problem sets and tests. 

These limitations point to directions for future work. Cross-disciplinary and cross-institutional studies could test where these patterns hold and how they vary across student populations and environments. Task design also warrants attention, since reading to familiarize oneself with a topic, as in our setting, may elicit different engagement than reading to construct an argument or prepare for a test. Mapping these boundaries will clarify when AI-supported reading support learning and when it erodes it.

\subsection{Limitations and Future Work}
\label{limitation and future work}

We recognize several others limitations of our study that are important to consider when interpreting our results besides the study context. 
First, we used students' prompts as proxies for cognitive engagement. While prompts externalize aspects of students' questioning and help-seeking behavior, they do not fully capture students' internal cognitive processes. Prior work has used similar prompt-coding approaches to characterize engagement patterns~\cite{ma2025scaffolding}, but future research could triangulate prompt analysis with complementary methods such as think-aloud protocols or comprehension assessments to validate the relationship between prompt types and underlying cognitive engagement. Second, we did not measure comprehension outcomes. The course graded reading assignments on a credit/no-credit basis: students received full credit for submitting their AI interaction logs regardless of depth or quality. This afforded ecological validity by capturing naturalistic AI use without performance pressure, but it precludes claims about whether observed engagement patterns actually affected learning outcomes. Third, our analysis centered on students' prompts rather than the AI responses they received. Because the quality and framing of AI output likely shape students' subsequent prompting and engagement, treating prompts in isolation is a limitation. We made this scoping decision for two reasons: students used different AI tools that we could not always identify, which made systematic comparison of response quality difficult, and rigorous evaluation of AI-generated text, whether automated or human, remains an open methodological challenge~\cite{lamsiyah2025trust, godbole2025verify}. Future work could examine how AI responses shape student behavior, for example through diary studies in which students record how particular responses affected their comprehension and how they assessed the responses' accuracy.
Fourth, the course may have influenced students' stated beliefs about prompting. Because the class explicitly taught prompt engineering, students may have stressed its importance during interviews, and we cannot separate this influence from public discourse or their own prior experience with AI.

\section{Conclusion}
We examined how college students use AI chatbots to support self-regulated reading over an eight-week period. Our findings reveal that students primarily used AI to outsource comprehension. Although students demonstrated natural cognitive progression within sessions, moving from comprehension toward reasoning, this progression was routinely truncated by minimal engagement. The intention-behavior gap we identified suggests that knowledge about effective AI use does not translate into practice: students articulated what good engagement looked like but optimized for efficiency instead. The novel pattern of reading \textit{through} AI rather than \textit{with} it, where students processed AI-generated summaries as primary material and triaged sections based on pre-existing interests, raises concerns about epistemic narrowing and the illusion of coverage.

These findings carry implications for both system design and pedagogy. General-purpose chatbots, which respond passively to user queries, may inadvertently reinforce shallow engagement. AI systems designed for reading support could instead scaffold sustained cognitive progression, surface unexplored content, and prompt metacognitive reflection. For educators, our results suggest that instruction alone is insufficient to shift engagement patterns; structural interventions such as minimum interaction requirements tied to learning objectives or comprehension assessments may be necessary. As AI tools become ubiquitous in academic contexts, understanding how students read with these systems, and designing them to support rather than circumvent deep engagement, becomes increasingly urgent.

\section*{Disclosure and Funding}
The authors report no competing interests to declare. No funding was received for this work.

\bibliographystyle{ACM-Reference-Format}
\bibliography{references}

\end{document}